\documentclass[12pt]{article}
\usepackage{amsmath,graphicx}

\begin{document}
\begin{center}
\large{\textbf{\textbf{ Spontaneous
 Generation of a Crystalline Ground State in a
Higher Derivative Theory
}}}\\
\end{center}
\begin{center}
Subir Ghosh\\
\vskip 1cm
Physics and Applied Mathematics Unit, Indian Statistical
Institute\\
203 B. T. Road, Kolkata 700108, India \\
\end{center}

{\textbf{Abstract}}\\
The possibility of Spontaneous Symmetry Breaking in momentum
space in a generic Lifshitz scalar
model -  a non-relativistic scalar field theory with higher spatial derivative
terms - has been studied. We show that the minimum energy state, the ground
state, has a lattice structure,
where the translation invariance of the continuum theory is reduced to a
discrete translation symmetry. The scale of translation symmetry breaking (or
induced lattice spacing) is
proportional to the inverse of the momentum of the condensate particle.
 The  crystalline ground state is stable under excitations below a certain
critical velocity. The small fluctuations
above the ground state can have a phonon like dispersion under suitable choice
of parameters.

 At the beginning we have discussed the effects of next to
nearest neighbour interaction terms
in a model of linear triatomic molecule depicted by a linear system of
 three  particles of same mass connected by identical springs. This model is
relevant since in the continuum limit
the next to nearest neighbour interaction terms generate higher (spatial)
derivative wave equation, the main topic of this paper.


  $email: sghosh@isical.ac.in $ ; FAX No. +(91)(33)25773026

\newpage
 (I) {\it{Introduction}}: Generating a continuum field theory as a limiting case
of a discrete set of coupled
 dynamical systems in the limit when the spacing between the systems goes to
zero is common. A text book example is
 the (nearest neighbor) coupled chain of harmonic oscillators that reduces to
the continuum elastic wave theory of sound in the limit
 when the oscillator spacing goes to zero. (See for example \cite{gold}.) 

On the other hand, quite
surprisingly, example of the complimentary phenomenon where a continuum field
theory can pass on to a 
theory with discrete degrees of freedom, (or more weakly where the continuous
nature of the ground state is replaced by 
a discrete structure), is comparatively rare. In the present paper we argue that
indeed this process can also be quite
common and can occure in any generic higher derivative theory. We provide an
explicit model, {\it{a higher derivative Lifshitz theory,
that yields a discrete ground state with a lattice structure}}. We also discuss the stability criteria of
such a ground state and properties of small excitations above the
ground state.

The necessity of higher derivative terms has been emphasized in a series of
papers by Polony1 and collaborators \cite{pol1,pol12,pol2,pol3}, in the context of
relativistic models where an instability in momentum space induces Spontaneous Symmetry Breaking, yielding
a discrete latticised ground state. Our present work is very
close in spirit to \cite{pol1} where the authors have revealed the rich phase
structure of a higher
derivative $\phi ^4$-theory, based on an inhomogeneous (periodic) ground state,
along with its one-loop renormalization.
Further, the model is renormalized to all orders in a perturbative framework and
is shown to be unitary in \cite{pol12}. It is well known that a relativistic 
higher derivative  field theory is plagued
with ghost excitation problem \cite{stel}. This issue is discussed in detail in
\cite{pol2} in the context of scalar QED with 
higher derivative terms where the gauge invariance plays a crucial role. This novel
phenomenon is extended to non-Abelian gauge
theory with higher derivatives in \cite{pol3}. The interesting aspect in these
works is that in situations where the
higher derivative term
in the kinetic part dominates, it can lead to a Spontaneous Symmetry Breaking
(SSB) in momentum space, resulting in a dynamical breaking of Lorentz
invariance. The ground state is
given by a condensate where the particles  carry a non-zero momentum
yielding an inhomogeneous spacetime dependent
ground state that breaks translation invariance. The (length) scale of
inhomogeneity is inversely proportional to the particle
momentum in the condensate.

As we have mentioned above, our work closely resembles \cite{pol1} with the
important distinction that we 
have considered a theory where Lorentz invariance is explicitly broken by the
higher derivative terms. Hence we
avoid the ghost problem (of higher derivative relativistic theories) altogether
and SSB of
Lorentz invariance is not an issue. No new excitations, Goldstone modes or
otherwise, are generated. In our model,
a non-relativistic theory is considered where the higher {\it{time}}
derivatives, responsible for the ghost, are dropped and
only spatial higher derivative terms are kept. In effective theories higher
derivative terms are 
naturally generated when one removes (or integrates out) some degrees of freeom. 
Such a theory, known as Lifshitz
theory \cite{lif}, is quite familiar in Condensed Matter Physics. But our main
interest lies in applying
this form of momentum space SSB in High Energy Physics, especially in Quantum
Gravity models.
In recent years Horava has proposed a conventional field theory for gravity
\cite{hor}  where spatial  higher derivative terms make the theory UV complete,
therby serving as a candidate of a 
Quantum Gravity model.   It is important to stress that a restricted form of
Horava
gravity model is essentially same as the model studied in the present paper, (as shown
by our earlier work \cite{dgh}). Hence the
results and conclusions reached here should be relevant not only in contexts of
Condensed Matter Physics but  in Horava theory of gravity
as well.

 In the present
paper we plan to study an interesting aspect of higher derivative generic
Lifshitz models which  so far have not been investigated. The higher derivative
terms can induce an instability leading to phase transition.
Furthermore,  the ground state (the lowest energy state) turns out to be
crystalline with
the lattice spacing dependent on the higher derivative coupling constant. Hence
the continuous translation
symmetry of the system is broken by the ground state that enjoys a discrete
translation symmetry only.

The above framework reminds us of  the celebrated Landau theory \cite{lan}
of liquid solid phase transition where Spontaneous Symmetry Breaking (SSB) leads
to a  state with less symmetry due to the structures present, the solid, from a
more uniform  and hence   symmetric
state, the liquid.  In a series of works, Alexander and  Mctague
\cite{alex} have shown in an essentially model independent way how crystalline
lattice
structure emerges from liquid. Rabinovici et.al, have utilised these ideas in
the context of String Theory compactification \cite{rab}. In the present case, a
similar thing happens: SSB generates a less
symmetric crystalline  vacuum condensate. The higher derivative
terms, a signature of the Lifshitz scalar model, is essential in inducing the
inhomogeneous condensate.
Very recently somewhat
similar ideas have been suggested by Wilczek and by Shapere and Wilczek in
\cite{wil}.

The paper is organized as follows: Section (II) deals with a mechanical
analogue toy model. In Section (III) we briefly
recapitulate SSB in a conventional scalar theory. In Section (IV) we present our
main results, the effects of SSB in momentum space in a generic Lifshitz model.
The paper ends with
conclusions and future outlook in Section (V).

(II) {\it{Mechanical analogue model for higher derivative wave equation}}: Below
we try to analyze the effect of higher
derivative term in a simple analogue
mechanical model. Although this toy model is not fully satisfactory but still
it indicates that the ground state
 of the system can be affected by such higher derivative (analogue) terms in a
way that is of interest for the present work. We
exploit the fact that higher derivative
terms can be simulated in a wave equation by {\it{next}} to nearest neighbour
interactions in the continuum limit of a lnear chain of mass points connectedc
by springs.

It is straightforward to generate the (elastic) wave equation from a chain of
mass points connected by
oscillators with nearest neighbour interactions, in the limit of the equilibrium
separation between the mass points, $a$, going to zero (see for example
\cite{gold}). With $\eta_i$ denoting displacement of the $i$-th mass point from
its equilibrium position and $\frac{d\eta}{dt}=\dot \eta ,~\frac{d\eta}{dx}=
\eta '$ the Lagrangian of the chain is
\begin{equation}
L=\frac{1}{2}\sum_i[m\dot \eta_i^2-\kappa
(\eta_{i+1}-\eta_i)^2)]$$$$=\frac{1}{2}\sum_i a[\frac{m}{a}\dot \eta_i^2-(\kappa
a) (\frac{\eta_{i+1}-\eta_i}{a})^2]
\label{v1}
\end{equation}
In the limit $a\rightarrow 0$ and $(\kappa a)\rightarrow Y$  identified as the
Young's modulus, it yields the equation of motion
\begin{equation}
\mu\ddot \eta -Y\eta ''=0,
\label{v2}
\end{equation}
where $(m/a)\rightarrow \mu$ is the mass per unit length and
$\frac{\eta_{i+1}-\eta_i}{a}\mid _{a\rightarrow 0}
\equiv \frac{\eta (x+a)-\eta (x)}{a}\mid _{a\rightarrow 0}=\frac{d\eta}{dx}$.
Now let us generalize
the potential term to include a next to nearest neighbour interaction term as
well. The resulting potential with
two coupling constants is,
\begin{equation}
L_{HD}=\frac{1}{2}\sum_i[m\dot \eta_i^2-\kappa_1
(\eta_{i+1}-\eta_i)^2)-\kappa_2\{(\eta_{i+1}-\eta_i)-(\eta_{i+2}-\eta_{i+1})\}^2
] $$$$=\frac{1}{2}\sum_i a[\frac{m}{a}\dot \eta_i^2-(\kappa_1
a) (\frac{\eta_{i+1}-\eta_i}{a})^2-(\kappa_2
a)(\frac{2\eta_{i+1}-\eta_{i+2}-\eta_i}{a})^2].
\label{v3}
\end{equation}
Similar considerations as above along with $(\kappa_1 a)\rightarrow Y,~(\kappa_2
a^3)\rightarrow F$ generates the higher derivative dynamics that we are
interested in (see also \cite{pol1}):
\begin{equation}
\mu\ddot \eta -Y\eta ''+F\eta ''''=0.
\label{v4}
\end{equation}

After establishing that $\kappa_2$-term induces a higher order interaction we
revert to a toy model, the
linear triatomic molecule, to study the effect of the $\kappa_2$ interaction
term. The system consists of three particles
of identical mass $m$, with identical springs attached between particles $1$ and
$2$ and between particles $2$ and $3$. Only vibrations along the line of the
molecule are being considered. The
interaction potential $V$ is given by,
\begin{equation}
V=\frac{\kappa_1}{2}[(\eta_2-\eta_1)^2)+(\eta_3-\eta_2)^2]+\frac{\kappa_2}{2}[
(\eta_2-\eta_1)-(\eta_3-\eta_2)]^2,
\label{v5}
\end{equation}
where $\eta _1$ refers to the displacement of particle $1$ from its equilibrium
position, etc.. Note that for $\kappa _2=0$ we recover the standard linear
triatomic molecule. The normal mode frequencies
are obtained as,
\begin{equation}
\omega_1=\frac{1}{\sqrt
m}[2\kappa_1+3\kappa_2+(\kappa_1^2+6\kappa_1\kappa_2+7\kappa_2^2)^{\frac{1}{2}}]
^{\frac{1}{2}}
,$$$$
\omega_2=\frac{1}{\sqrt
m}[2\kappa_1+3\kappa_2-(\kappa_1^2+6\kappa_1\kappa_2+7\kappa_2^2)^{\frac{1}{2}}]
^{\frac{1}{2}}.
\label{v6}
\end{equation}
Let us try to find the (static)
ground state of the system by minimizing the potential $V$, subject to the
condition $\eta_1+\eta_2+\eta_3=0$ indicating
that only vibrations are considered with the centre of mass fixed at origin. The
conditions are,
$$\frac{\partial (V+\lambda (\eta_1+\eta_2+\eta_3))}{\partial \eta_i}=0;
~i=1,2,3;~\frac{\partial (V+\lambda (\eta_1+\eta_2+\eta_3))}{\partial \lambda}=0
$$ where $\lambda$ is a Lagrange multipliar. For $\kappa_2=0$ the set
of equations,
\begin{equation}
\kappa_1 (2\eta_2-\eta_1-\eta_3)+\lambda =0;~\kappa_1 (\eta_2-\eta_1)+\lambda
=0,$$$$
\kappa_1 (\eta_3-\eta_2) +\lambda =0;~\eta_1+\eta_2+\eta_3 =0,
\label{v7}
\end{equation}
have the solution $\eta_1=\eta_2=\eta_3=0$. Hence there is a unique ground state
with all the particles at their
equilibrium position.

However, for non-zero $\kappa_1$ and $\kappa_2$ the set of equations,
\begin{equation}
\kappa_1 (\eta_2-\eta_1)+\kappa_2  (2\eta_2-\eta_1-\eta_3)+\lambda =0;~(\kappa_1
+2\kappa_2) (2\eta_2-\eta_1-\eta_3)+\lambda
=0,$$$$
\kappa_1 (\eta_3-\eta_2)-\kappa_2(2\eta_2-\eta_1-\eta_3)+\lambda
=0;~\eta_1+\eta_2+\eta_3 =0,
\label{v8}
\end{equation}
yield $\eta_1=\eta_3=-2\eta_2$ for $\kappa_1+2\kappa_2=0$. This shows that there
can appear a one parameter family of ground states of the system with infinitely
many
possibilities. Without the condition $\kappa_1+2\kappa_2=0$ we again obtain
$\eta_1=\eta_2=\eta_3=0$. Unfortunately the condition on $\kappa$-s lead to
complex normal mode frequencies. It is plausible that more complex systems with
larger number of degrees of freedom can lead stable oscillatory modes with a
family of
ground states as above. But we have
succeeded in showing that the ground state can indeed be influenced by  effect
of analogue higher derivative interaction terms. Fortunately, in the higher
derivative field theory model discussed below (that is our true interest), these
pathologies are absent.

(III) {\it{Spontaneous Symmetry Breaking in normal scalar theory}}: Let us
briefly recall SSB in a conventional scalar theory. The Lagrangian
\begin{equation}
{\it{L}}=\frac{(\dot \phi )^2}{2}-(a\frac{ (\phi ') ^2}{2}+c\frac{ \phi
^2}{2}+\lambda \frac{ \phi ^4}{2}),
\label{l}
\end{equation}
leads to the energy,
\begin{equation}
E=\int~ dx~(\frac{(\dot \phi )^2}{2}+(a\frac{ (\phi ') ^2}{2}+c\frac{ \phi
^2}{2}+\lambda \frac{ \phi ^4}{2}).
\label{e}
\end{equation}
The minimum energy ground state will obviously correspond to a spacetime
constant $\phi$ obtained from the solution of $(\partial E)/(\partial \phi )=0$
yielding $\phi_S=0,~\phi_{BS}=\pm {\sqrt{(-c)/2\lambda }}$ with the
corresponding
energies $E_S=0,~E_{BS}=-c^2/(8\lambda )$. $S,BS$ stand for Symmetric and Broken
Symmetric phases respectively. For the ground state to exist $c<0,~\lambda >
0$ so that $E_{BS}<E_S$ indicating that the symmetry broken phase has a lower
energy. Incidentally, the symmetry in question that is broken is the
$\phi\rightarrow -\phi $ reflection symmetry that either of the ground states
$\phi_{BS}$ fail to preserve.

Several  well known features are worth pointing out for contrasting and
comparing with the SSB phenomenon we  are going to present later. {\it{(i)}} For
the
Lorentz invariant scalar theory one needs to have $a=1$ and for a conventional
massive theory $c>1$. However for  SSB to occur in the relativistic theory with
positive $\lambda $ we need $c<0$. But this not a problem since after shifting
$\phi$ appropriately for one of the BS ground states the resulting scalar gets
the correct sign mass term. {\it{(ii)}} Both $\phi_S ,\phi_{BS}$ are
solutions of the equation of motion. {\it{(ii)}} From a momentum space point of
view with $\phi (x)=\int dk \varphi (k) exp (ikx)$ the S and BS ground states
with {\it{constant}} $\phi $ indicate
that $\varphi (k)$ has a maximum at $k=0$ only that is $\varphi (k) \sim \delta
(k)$. This suggests that, {\it{to construct a theory with   a variable ground
state}} $\phi (x)$ {\it{we must
look for a model where the Fourier transform}} $\varphi (k)$ {\it{has a peak
at some non-zero momentum}}
$k=\bar k$. Keeping in mind the structure of $\phi$-terms in (\ref{e}), for SSB
to occur in momentum space we will need at least a fourth order (space)
derivative term that is quadratic
in $\phi$.  In the rest of the paper we have precisely constructed and studied
such a model in the context of Lifshitz scalar theory.

(IV) {\it{Spontaneous Symmetry Breaking in Lifshitz scalar theory}}:
After this brief recapitulation of SSB in  conventional scalar $\phi $ let us
move  to the arena of Lifshitz scalar field theory. We posit a generic Lifshitz
Lagrangian as,
\begin{equation}
{\it{L_{Lif}}}=\frac{(\dot \phi )^2}{2}-(a\frac{ (\phi ') ^2}{2}+b\frac{(\phi
'')^2}{2}+c\frac{ \phi ^2}{2}+\lambda \frac{ \phi ^4}{2}),
\label{ls}
\end{equation}
where the $b$-term represents a higher spatial derivative term. Clearly we have
sacrificed Lorentz invariance. Again we look for the ground state. The static
energy is
\begin{equation}
E_{Lif}=\int~ dx~(a\frac{ (\phi ') ^2}{2}+b\frac{(\phi '')^2}{2}+c\frac{ \phi
^2}{2}+\lambda \frac{ \phi ^4}{2}).
\label{es}
\end{equation}
Once again a constant $\phi =\tilde \varphi$ can be a possible ground state with
\begin{equation}
\tilde E=\int~ dx~(c\frac{ \phi ^2}{2}+\lambda \frac{ \phi ^4}{2}).
\label{1}
\end{equation}
Minimizing $\tilde E$ leads to a normal phase with
$\tilde \varphi_S =0,~ \tilde E_S=0 $ or a broken symmetry  phase $\tilde
\varphi_{BS} ={\sqrt{-c/2\lambda }},~ \tilde E_{BS}=-c^2/(8\lambda ) $. Since
$\lambda$ has to be positive for the ground state to exist $c<0$ for the broken
symmetry phase. As far as constant-$\phi$ ground state is concerned this is same
as the analysis below (\ref{e})
with the $b$-term having no effect but indeed, this is not what we are after.

Now we discuss the more interesting possibility, that of a space dependent
$\phi_{BS}(x)$ ground state. We introduce a Fourier transform: $\phi (x)=\int dk
\varphi (k)
exp(ikx)$. In the conventional case without the higher derivative term one has
$\phi=constant $ as the minimum energy state with non-zero constant value for
SSB. A constant $\phi$ yields
a vanishing kinetic energy part. This is compatible with $k=0$ in the Fourier
transform $\varphi (k)$ leading to a constant $\phi$. But with the higher
derivative term present there can be
more options for minimizing the energy. One possibility, as we show below, is a
non-zero $\bar k$ value leading to
a space dependent $\phi(x)$ through the Fourier transform $\varphi(\bar k)$.
The derivative part of the energy in momentum space gives
\begin{equation}
E_{Lif}(derivative)=\int~dk (a\frac{k^2}{2}+b\frac{k^4}{2})\varphi (k)\varphi
(-k)=2\int~dk (a\frac{k^2}{2}+b\frac{k^4}{2})(\varphi (k))^2,
\label{ee}
\end{equation}
where in the last step we have assumed $\varphi (k)=\varphi (-k)$. Clearly for
ground state to exist $b>0$ and minimizing with respect to $k$ we find
\begin{equation}
k_N=0 ~for~a>0,~b>0,~\bar E(k_N)=\tilde E_S=-\frac{c^2}{8\lambda} ,$$$$
k_{BS}\equiv\bar k={\sqrt{\frac{-a}{2b}}}
~for~a<0,~b>0,
\label{k}
\end{equation}
which in turn lead to $\varphi_S (k)=\bar\varphi \delta(k)$ and $\varphi_{BS}
(k)=\bar\varphi \delta(k-\bar k)$.
These indicate a constant $\phi_S=\bar\varphi $ in the first case (that we have
already studied above) and a spatially varying $\phi_{BS} (x) =\bar \varphi
exp(i\bar k x)$ in the second case which is of interest to us. Since we have
already assumed $\varphi (k)=\varphi (-k)$, let us consider the broken symmetry
condensate to be,
 \begin{equation}
\phi_{BS} (x) =\bar \varphi cos(\bar k x).
\label{g}
\end{equation}
 We consider $\phi_{BS} (x)$ to be a candidate solution for the ground state in
the broken symmetry phase of the Lifshitz model. We still need to find $\bar
\varphi $. We compute this by putting back $\phi_{BS} (x)$ of (\ref{g}) in to
the energy (\ref{es}) and minimizing it. The energy becomes,
\begin{equation}
\bar E=\int ~dx~\frac{1}{2}[(a\bar k^2+b\bar k^4 +c)(\bar \varphi )^2 cos^2(\bar
kx)+\lambda (\bar \varphi )^4 cos^4(\bar kx)]$$$$
=\int ~dx~[(-\frac{a^2}{8b}+\frac{c}{2})(\bar \varphi )^2 cos^2(\bar
kx)+\frac{\lambda}{2}(\bar \varphi )^4 cos^4(\bar kx)].
\label{egr}
\end{equation}
We consider an approximation {\footnote{Explicitly $\int _{-L}^L
cos^2(\bar kx)dx=L+\frac{sin(2\bar kL)}{2\bar k},~\int _{-L}^L
cos^4(\bar kx)dx=\frac{1}{4}(3L+2\frac{sin(2\bar kL)}{\bar k}+\frac{sin(4\bar
kL)}{4\bar k})$. Hence to $O(\bar kL)$ the integrals are equal to $2L$.}}
$cos^2(\bar
kx)\approx cos^4(\bar kx)$ which is reasonable for small $\bar k x$ and obtain ,
\begin{equation}
\bar E=\int ~dx~[(-\frac{a^2}{8b}+\frac{c}{2})(\bar \varphi
)^2+\frac{\lambda}{2}(\bar \varphi )^4)]
cos^2(\bar kx).
\label{ebar}
\end{equation}
Minimizing
the functional with respect to  $\bar\varphi$
\begin{equation}
(c-\frac{a^2}{4b})\bar \varphi +2\lambda \bar \varphi ^3 =0
\label{phi}
\end{equation}
the solutions of the resulting equation  are
\begin{equation}
\bar \varphi_S =0,~ \bar
\varphi_{BS}={\sqrt{\frac{1}{2\lambda}(\frac{a^2}{4b}-c)}}.
\label{sol}
\end{equation}
For the non-zero solution, since $\lambda >0$, the parameters have to satisfy
the inequality $\frac{a^2}{4b}-c >0$. Note that, although not essential, we can
restrict to $c<0$ to include SSB with a constant $\phi $ ground state (see below
(\ref{1}), $\tilde
\varphi_{BS} ={\sqrt{-c/2\lambda }},~ \tilde E_{BS}=-c^2/(8\lambda ) $, for
which
the inequality is automatically satisfied. Thus, for the Lifshitz scalar a
possible
broken symmetry space dependent solution is
\begin{equation}
\phi_{BS} (x)={\sqrt{\frac{1}{2\lambda}(\frac{a^2}{4b}-c)}} ~cos
({\sqrt{\frac{-a}{2b}}} x).
\label{con}
\end{equation}
This is our principal result. Since the state has a non-zero momentum $\bar k$
it is termed as
a flowing state following \cite{wen}.

Before proceeding further let us summarize. For the generic Lifshitz model we
can have three possible ground states: the symmetric $\phi =0$ phase, the broken
symmetry constant $\phi $ phase and
finally the broken symmetry variable $\phi $ phase. We are interested in the
last one.

In rest of the paper we establish the
claim of $\phi_{BS}(x)$ being a viable ground state by ensuring three crucial
points:
{\it{(i)}} $\bar\phi_{BS} (x)$ is a solution of the equation of
motion. This condition leads to the crystalline ground state. {\it{(ii)}} Energy
of
the space dependent  ground state $\phi _{BS}$ can be lower than the energy of
the constant $\phi_S$ ground state otherwise the $\phi_{BS}$ ground state will
not be selected by the system. {\it{(iii)}} The ground state with non-zero
momentum $\bar k$ will be stable against decay
into small fluctuations.

{\it{(i)}} The equation of motion is derived as,
\begin{equation}
\ddot \phi(x,t) -a\phi ''(x,t)+b\phi ''''(x,t)+c\phi(x,t) +2\lambda \phi
^3(x,t)
=0.
\label{eqm}
\end{equation}
Substitution of $\phi_{BS} (x)$ from (\ref{con}) in (\ref{eqm}) yields
\begin{equation}
\frac{1}{\sqrt {2\lambda }}(\frac{a^2}{4b}-c)^{3/2}cos (\bar k
x)(cos^2(\bar k x)-1)=0.
\label{k1}
\end{equation}
Since the other factors are non-vanishing, (note that $cos(\bar kx)=0\rightarrow
\phi =0$ state),  a possible condition emerges on the space coordinate itself:
\begin{equation}
cos^2(\bar k x)-1=0 ~~\rightarrow ~~cos (\bar k x)=\pm 1,~ x_n=\frac{\pi}{\bar
k}n=\pi{\sqrt{\frac{2b}{-a}}}n,~n=0,1,2,...
\label{n}
\end{equation}
This shows that the ground state is lattice like with non-zero values of
$\phi_{BS}(x_n)$, the
lattice spacing being $\Delta x=\pi /\bar k=\pi{\sqrt{2b/(-a)}}$. As advertized
before, the scale of 
translation symmetry breaking, in the form of lattice spacing $\Delta x$ is
inversely proportional to $\bar k$, the momentum of 
the condensate particles \cite{pol1}.  This is the
most important outcome of our result (\ref{con}), {\it{the crystalline
condensate}}.
Clearly
a necessary condition for this to occur is the presence of the higher derivative
$b$-term in (\ref{ls}). Hence at length scales $\sim~(\bar k)^{-1}$ the ground
state will consist of alternate
values of $\pm \bar\varphi $ at the discrete points $x_n${\footnote{The ground atate has a formal similarity
with antiferromagnetic Neel state. Something similar also occured in \cite{pol1}.}}. However at larger
length (or
low energy) the ground state
will have  a smoother wave-like periodic behavior. \\
{\it{(ii)}} The energy of the condensate ground state is now computed from
(\ref{ebar})
\begin{equation}
\bar E(\bar k)=-\frac{c^2}{8\lambda}(\frac{a^2}{4bc}-1)^2=-\mid \bar E(k_N)\mid
(\frac{a^2}{4bc}-1)^2.
\label{ee1}
\end{equation}
Since we have already restricted to $c<0$, it is  true that $\mid \bar E(\bar
k)\mid >\mid \bar E(k_N)\mid $ and hence the variable
$\phi_{BS}(x)$  energy is lower that the constant $\tilde\phi$ ground state.
Alternatively if
we do not restrict $c$ to be negative, we know $\frac{a^2}{4bc}-1>0$ (see below
(\ref{sol}) and hence to satisfy the above we just  require
$\frac{a^2}{4bc}-1>1$.
\begin{figure}[htb]
{\centerline{\includegraphics[width=9cm, height=6cm] {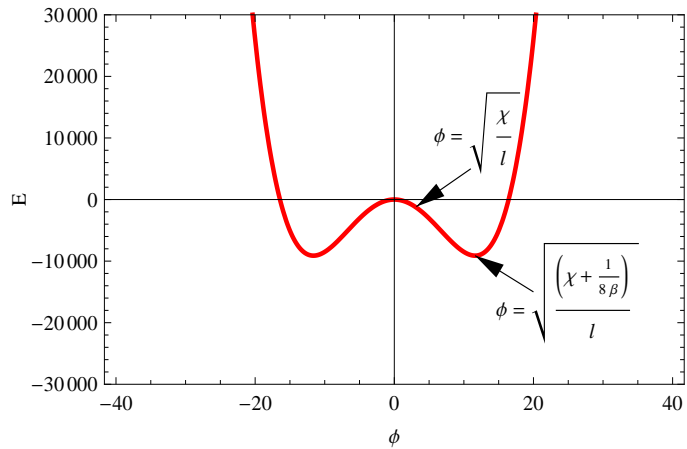}}}
\caption{{\it{The energy profile shows that the condensate ground state can be lower
than the constant $\varphi $
ground state. We have scaled $c\equiv a\chi, b\equiv a\beta, \lambda \equiv
al$.}}} \label{fig1}
\end{figure}\\
{\it{(iii)}} We consider a small fluctuation above the flowing ground state,
$\phi (x,t)=\bar \varphi cos(\bar kx)+\delta \phi (x,t)$ and the Lagrangian,
quadratic in $\delta \phi (x,t)$ becomes,
\begin{equation}
{\cal{L}}(O(\delta\phi )^2)=(\delta \dot \phi )^2-[a(\delta \phi')^2+b(\delta
\phi
'')^2+c(\delta\phi )^2+6\lambda\bar \varphi ^2(\delta\phi )^2]
\label{p}
\end{equation}
In the $\lambda$-term above, we have replaced $\phi$ by $\bar \varphi$ as it is
already  $O(\delta\phi )^2$. To derive the minimal stability condition we
introduce a time-independent Fourier transform $\delta\phi =\sum _k
\delta\phi_kexp(ikx)$ and find,
\begin{equation}
E(\delta\phi )=\sum _k (ak^2+bk^4+c+6\lambda \bar \varphi ^2)\phi_{-k}\phi_k.
\label{ek}
\end{equation}
The energy will be positive if the function within the bracket is positive.
Since $bk^4$ is positive definite, the positivity condition reduces to
\begin{equation}
ak^2+3(\frac{a^2}{4b}-\frac{2c}{3})>0 ~\rightarrow ~-\mid a \mid k^2
+3(\frac{a^2}{4b}-\frac{2c}{3})>0,
\label{pos}
\end{equation}
and finally
\begin{equation}
k^2<\frac{3}{4}\frac{\mid a\mid}{b}-\frac{2c}{\mid a\mid}~ \rightarrow
~k^2<\frac{3}{2}\bar k^2-\frac{2c}{\mid a\mid}.
\label{pos1}
\end{equation}
Again for $c<0$ this boils down to $k^2<\frac{3}{4}\bar k^2+\frac{2\mid
c\mid}{\mid a}$. Physically this means that so long as the excitation velocities
are below (of the order of) $\bar k$ the excitation energies are positive and
thus they are not capable of reducing the energy of the flowing ground state
thus
making the latter
stable \cite{wen}.

Next, following \cite{wen}, we compute the spectra of the small fluctuations
above the condensate
ground state $\phi_{BS}(x,t)=\bar\phi_{BS} (x)+\delta\phi_{BS}(x,t)$. The
equation of
motion gives
\begin{equation}
\ddot{(\delta\phi_{BS})}-a(-\bar k^2\bar\varphi +(\delta\phi_{BS})'')+b(\bar
k^4\bar\varphi +(\delta\phi_{BS})'''')$$$$
+c(\bar\varphi +\delta\phi_{BS})+2\lambda \bar\varphi ^2(\bar\varphi
+3\delta\phi_{BS})=0.
\label{fl}
\end{equation}
In the above we have used the condition (\ref{n}). By substituting $\bar\varphi$
and $\bar k$ and cancelling out
the inhomogeneous terms, we obtain
\begin{equation}
\ddot{(\delta \phi_{BS} ) }-a(\delta \phi_{BS} )''+b(\delta\phi_{BS}
)''''+(c+6\lambda \bar \varphi ^2)(\delta \phi_{BS} )=0.
\label{del}
\end{equation}
Consider a wave solution of the form
\begin{equation}
\delta \phi_S(x,t)=Ae^{-i\omega_kt+ikx}+Be^{i\omega_kt-ikx}.
\label{sp}
\end{equation}
The energy spectra is
\begin{equation}
\omega_k=\pm{\sqrt{ak^2+bk^4+(\frac{3a^2}{4b}-2c)}}.
\label{en}
\end{equation}
For small $k$ we can drop $bk^4$ so long as $k^2<\frac{3}{2}\bar k^2+\frac{2\mid
c\mid}{\mid a\mid}$, (since $a$ is negative). The  modes, for small $k$, will be
gapless and phonon-like provided
$\frac{3a^2}{4b}-2c =0$.

(V) {\it{Conclusion and future prospects}}: Let us summarize our work. It is well known that 
nearest
neighbor interactions in a linear chain of mass points, connected by springs,
generate the elastic wave theory in continuum limit. We have shown in a toy model that, in an analogous way,
 next to  nearest
neighbor interactions in the same system can generate higher (spatial)
derivative wave equation. We have also demonstrated how the next to  nearest
neighbor interactions in a linear triatomic molecule can yield a spatially
inhomogeneous ground
state.

Subsequently we come to the main
body of our work. We have considered exhaustively various
possibilities of Spontaneous Symmetry Breaking in momentum space that can be
induced by higher spatial
derivative terms in a Lifshitz scalar model.
We have shown that a flowing ground state condensate, with a crystalline
structure is certainly possible that can have the
lowest ground state energy with respect to other conventional ground state of
pure vacuum or having a spatially homogeneous condensate. The stability of this
novel
ground state against decay into
small disturbances is discussed. Spectra of fluctuations lying above the flowing
condensate have an unconventional spectrum. As we have
discussed above, the ground state
will look like a discrete set of spin half particles with positive/negative
values of the condensate amplitude $\pm \bar\varphi $ at the alternate discrete
 lattice points. At long wavelength the ground state will appear to be wavelike.

It is intriguing to consider this work in the context of Horava-Lifshitz
gravity, where
the the higher derivative terms tend to improve
the short distance behavior of gravitation whereas in the long wavelength regime
these Lorentz non-invariant terms will not be
important and Einstein gravity will prevail. This is a new perspective where the
Lorentz covariant Einstein gravity appears as an effective
low energy theory. From our earlier work we know that the  Horava-Lifshitz
gravity theory, under certain restrictions, can be
reduced to the generic Lifshitz form that we have
analyzed here. In this case the Lifshitz scalar field $\phi$ will be replaced by
the metric  field and it is worth speculating
that the crystalline ground state condensate obtained here might be interpreted
as a
"space-crystal". This is relevant because,
from quite  general perspectives, it is  expected that space(time) might be
discrete at
Planck scales so the present work can be extended to higher dimensions (which is
straightforward) to generate a discrete "space-crystal" or even a
"spacetime-crystal".
\vskip .2cm
{\it{Acknowledgments}}: I thank Professor Chetan Nayek and
Professor Eliezer Rabinovici for helpful suggestions at an early stage of the
work. Also I thank Sudipta Das for
discussions and Shom Shankar Bhattacharya for solving the generalized triatomic
molecular
dynamics problem. I am grateful to the anonymous referee for constructive
suggestions.

\end{document}